\documentclass[10pt]{article}
\usepackage[OE]{express}
\usepackage{subfigure}
\usepackage{graphicx}
\usepackage{epstopdf}
\usepackage{cite}
\usepackage{amssymb}

\newcommand{\ket}[2][]{{|#2\rangle_{#1}}}
\newcommand{\bra}[2][]{{}_{#1}\langle #2|}

\renewcommand\maketitle{}

\begin{document}
\title{Quantum fingerprinting using two-photon interference}

\author{Micha\l{} Jachura,\authormark{1,*} Micha\l{} Lipka,\authormark{1} Marcin Jarzyna,\authormark{2} and Konrad~Banaszek\authormark{2}}

\address{\authormark{1}Faculty of Physics, University of Warsaw, Pasteura 5, 02-093 Warszawa, Poland\\
\authormark{2}Centre of New Technologies, University of Warsaw, Banacha 2, 02-097, Warszawa, Poland\\
}

\email{\authormark{*}michal.jachura@fuw.edu.pl} 



\begin{abstract}
We present a quantum fingerprinting protocol relying on two-photon interference which does not require a shared phase reference between the parties preparing optical signals carrying data fingerprints. We show that the scaling of the protocol, in terms of transmittable classical information, is analogous to the recently proposed and demonstrated scheme based on coherent pulses and first-order interference, offering comparable advantage over classical fingerprinting protocols without access to shared prior randomness. We analyze the protocol taking into account non-Poissonian photon statistics of optical signals and a variety of imperfections, such as transmission losses, dark counts, and residual distinguishability. The impact of these effects on the protocol performance is quantified with the help of Chernoff information.
\end{abstract}


\section{Introduction}

The use of quantum mechanical superposition states allows one to
reduce the communication complexity in certain tasks involving
distributed information processing. A seminal example has been provided
by Buhrman {\em et al.} with the quantum fingerprinting protocol \cite{Buhrman2001}, where a
Referee is tasked with deciding whether bit strings in possession of two
separate parties, Alice and Bob, are identical or different.
If Alice and Bob can sent to Referee quantum systems prepared in
superposition states rather than orthogonal ones that can carry
only classical information, the resource scaling changes to
logarithmic from square-root in terms of the number of transmitted qubits. The classical bound is based on an assumption that Alice and Bob have no access to shared prior randomness.
Interestingly, the quantum fingerprinting protocol has a feasible
optical implementation using coherent states and passive linear optics as recently
proposed \cite{Luthenhaus2014,Kumar} and experimentally demonstrated over standard telecom fiber links \cite{Xu,Guan}. Comparing bit strings encoded in optical signals is also used as a primitive in other quantum protocols, such as quantum digital signatures \cite{Clarke,Collins}.
Note that the coherent-state fingerprinting protocol has a quantum mechanical single-particle counterpart with field amplitudes replaced by probability amplitudes \cite{Massar}.

The coherent-state implementation of the quantum fingerprinting
protocol relies on phase coherence between Alice and Bob that
enables the Referee to obtain the required information using
first-order interference between received coherent states. Although
substantial effort has been dedicated to develop stable phase-locked light sources
for coherent optical communication \cite{Kitayama}, they may be unavailable or
impractical in some specialized scenarios. A possible remedy to this problem is to resort to
second-order interference \cite{Mandel}, which is insensitive to the global phase
between the incoming beams. In this paper we present and discuss theoretically a quantum fingerprinting protocol, which does not require a shared phase reference between the communicating parties. It is based on two-photon interference also known as the Hong-Ou-Mandel effect \cite{HOMI} which can serve as a linear-optics implementation of the swap test \cite{Garcia}.
The principle of quantum fingerprinting has been previously demonstrated in experiments with pairs of qubits \cite{Horn,Du}.
Here we analyze a scalable version of the protocol based on mapping the bit strings held by Alice and Bob onto sequences of binary $\pm$ phases that are used to modulate single photons prepared as superposition states of multiple time bins. We show that the advantage over classical protocols without shared randomness matches that for the coherent state scenario despite lack of the phase reference.

Further, we analyze the impact of imperfections, such as residual distinguishability between interfering photons, non-unit detection efficiency which can also include losses and probabilistic single-photon generation, as well as the presence of multiphoton components in the radiation employed in the fingerprinting protocol. The last imperfection is characterized with the help of the normalized second-order intensity correlation function $g^{(2)}$ for a zero delay time \cite{Mandel}. This model can describe quantum fingerprinting using either single photons, corresponding to $g^{(2)} \approx 0$, or phase-averaged coherent states with Poissonian statistics for which $g^{(2)} = 1$, in the regime when the probability of detecting a photon in a single experimental run is much less than one. Such a situation can occur for low transmission of optical channels used by Alice and Bob to send optical signals to the Referee. The performance of a realistic implementation of the two-photon quantum fingerprinting protocol is quantified using the concept of Chernoff information \cite{CoverThomas}, which is a convenient information theoretic tool to characterize the attainable error probability in the asymptotic limit of a large number of experimental runs.

This paper is organized as follows. In Sec.~\ref{Sec:TwoPhotonProtocol} we review quantum fingerprinting with coherent states and introduce the two-photon protocol. The construction of the error-correcting code required for the two-photon protocol is presented in Sec.~\ref{Sec:ECC}. The advantage in terms of transmittable information compared to classical protocols is also discussed there. In Sec.~\ref{Sec:realistic} we describe a model that includes common experimental imperfections. The effect of using light sources with different photon statistics is analyzed using Chernoff information in Sec.~\ref{Sec:Chernoff}. Finally, Sec.~\ref{Sec:Conclusions} concludes the paper.

\section{Fingerprinting using optical interference}
\label{Sec:TwoPhotonProtocol}

\begin{figure}
\includegraphics[scale=0.8]{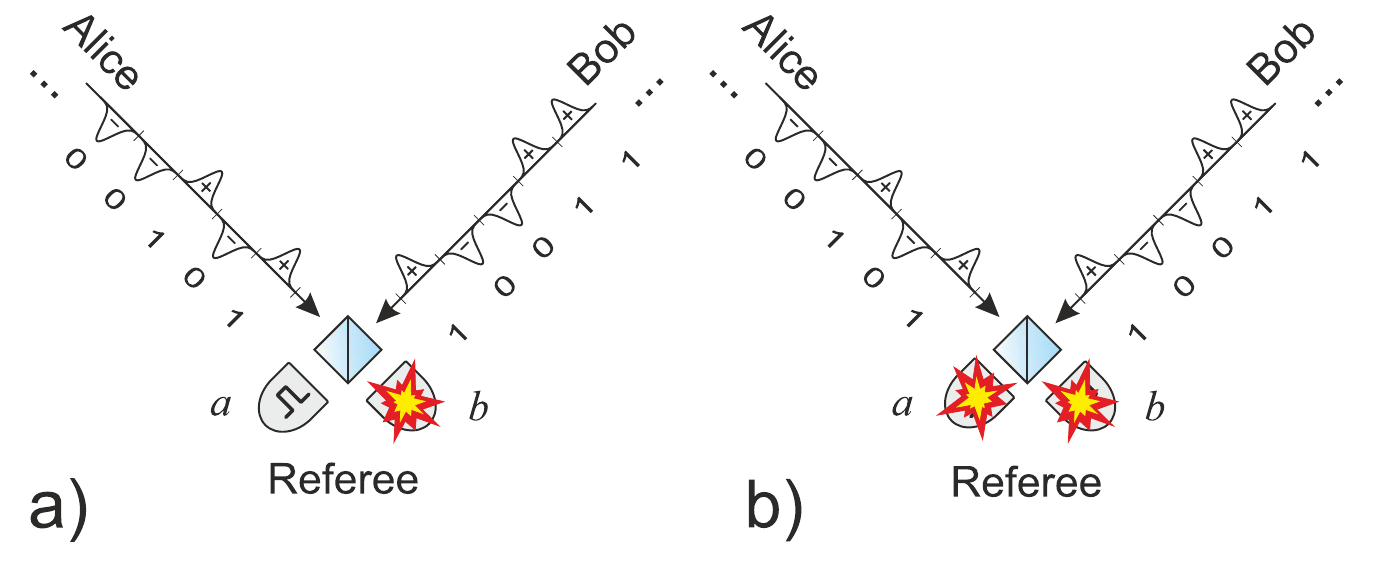}\centering\caption{Quantum fingerprinting protocols using (a) coherent states and (b) photon pairs. Alice and Bob map error correcting codewords onto
$+/-$ phase modulation patterns for optical signals sent to the Referee, who combines them on a 50/50 beam splitter. (a) In the case of first order interference, a click on the detector monitoring the output port $b$ of the beam splitter, where field amplitudes are subtracted, implies that the codewords were different. (b) The same conclusion can be drawn in the two-photon protocol when a coincidence between the two detectors is registered.}
\label{fig:Scheme}
\end{figure}

The basic idea of quantum optical fingerprinting \cite{Luthenhaus2014} is to map $n$-bit strings ${\tt x}$ and ${\tt y}$ held by Alice and Bob respectively onto sequences of $m$ optical pulses that are subsequently sent to the Referee for comparison. In the simplest scenario Alice and Bob modulate individual pulses with two opposite phases $\pm$ that are chosen according to $m$-bit long binary error correcting codewords ${\tt E}({\tt x})$ and ${\tt E}({\tt y})$ for the input strings. The purpose of the error correcting code is to enhance the differences between input strings when ${\tt x} \neq {\tt y}$ and thus to augment the response obtained at the detection stage. The relevant characteristics of the error correcting code ${\tt E}$ will be discussed in Sec.~\ref{Sec:ECC}. If $\hat{a}_i$ and $\hat{b}_i$, $i=0,\ldots,m-1$, denote annihilation operators for modes corresponding to individual pulses on Alice's and Bob's sides, such modulation corresponds to preparing states of the electromagnetic field  over all $m$ time bins in modes $\hat{a}_{\tt x}$ and $\hat{b}_{\tt y}$ that are given by linear combinations:
\begin{equation}
\hat{a}_{\tt x} = \frac{1}{\sqrt{m}}\sum_{i=0}^{m-1} (-1)^{{\tt E}_i({\tt x})} \hat{a}_i ,
\qquad \hat{b}_{\tt y} = \frac{1}{\sqrt{m}}\sum_{i=0}^{m-1} (-1)^{{\tt E}_i({\tt y})} \hat{b}_i.
\end{equation}
In the standard coherent-state protocol \cite{Luthenhaus2014} the states of optical signals prepared by Alice and Bob can be written as
\begin{equation}
 \ket[A]{\alpha_{\tt x}} = \exp(\alpha\hat{a}_{\tt x}^\dagger - \alpha^\ast \hat{a}_{\tt x}) \ket[A]{0}, \qquad  \ket[B]{\beta_{\tt y}} = \exp(\beta\hat{b}_{\tt y}^\dagger - \beta^\ast \hat{b}_{\tt y}) \ket[B]{0},
\label{Eq:InputState}
\end{equation}
where $\alpha$ and $\beta$ are the complex amplitudes of the sequences prepared by Alice and Bob, and $\ket[A]{0}$ and $\ket[B]{0}$ denote vacuum states on Alice's and Bob's sides respectively. Let the complex amplitudes be equal and have the same phase, $\alpha= \beta=\sqrt{\bar{n}}$, where $\bar{n}$ is the mean total photon number in each of the sequences. The signals received from Alice and Bob are interfered by the Referee on a balanced beam splitter as shown in Fig.~\ref{fig:Scheme}(a). This corresponds to the standard linear transformation of the annihilation operators for individual pulses \cite{GerryKnight}:
\begin{equation}
\hat{a}_i \rightarrow (\hat{a}_i + \hat{b}_i)/\sqrt{2}, \qquad \hat{b}_i \rightarrow (\hat{a}_i - \hat{b}_i)/\sqrt{2}.
\label{Eq:BeamSplitterTransformation}
\end{equation}
In the case of identical strings, when ${\tt x} = {\tt y}$, completely destructive interference occurs at the beam splitter output port $b$ corresponding to the family $\hat{b}_i$ of annihilation operators in the representation of output modes. For unequal strings, when ${\tt x} \neq {\tt y}$, the field amplitude at this port is non-zero. A straightforward calculation presented in Appendix yields the probability of registering at least one count at this port equal to $1-\exp[-\bar{n}(1-\text{Re} {\cal V})]$, where
\begin{equation}
{\cal V} = \frac{1}{m} \sum_{i=0}^{m-1} (-1)^{{\tt E}_i({\tt x}) + {\tt E}_i({\tt y})}
\label{Eq:VisibilityDef}
\end{equation}
is the interference visibility. The right hand side of the above expression can be generalized in a straightforward manner to the case when the phases of individual pulses are taken from a constellation larger than the binary set $+,-$. The protocol gives an erroneous answer when the interference of optical signals encoding different strings does not generate any click at the output port $b$.
The  probability of such an event is given by $ \exp[-\bar{n}(1-\text{Re} {\cal V})] $, which exhibits exponential scaling with the mean photon number $\bar{n}$.
In many physical scenarios, sharing a common phase reference between Alice and Bob or maintaining the phase stability during the realization of the protocol may be difficult if not impossible. This requirement could be especially demanding in free-space optical communication used e.g.\ in earth-satellite links or in signal transmission between moving objects. If the global phase between sequences prepared by Alice and Bob is random, a detection event on a single output port can no longer serve as an unambiguous signature of different strings ${\tt x} \neq {\tt y}$. However, in this case one can resort to the Hong-Ou-Mandel effect \cite{HOMI} in two-photon interference which is invariant with respect to a global phase shift.

Suppose that Alice and Bob prepare single photons in modes $\hat{a}_{\tt x}$ and $\hat{b}_{\tt y}$:
\begin{align}
\ket[A]{1_{\tt x}} & = \hat{a}^\dagger_{\tt x} \ket[A]{0} = \frac{1}{\sqrt{m}}\sum_{i=0}^{m-1} (-1)^{{\tt E}_i({\tt x})} \hat{a}_i^\dagger \ket[A]{0} \nonumber \\
\ket[B]{1_{\tt y}} & = \hat{b}^\dagger_{\tt y} \ket[B]{0} = \frac{1}{\sqrt{m}}\sum_{i=0}^{m-1} (-1)^{{\tt E}_i({\tt y})} \hat{b}_i^\dagger \ket[B]{0},
\end{align}
where the strings are encoded into binary $\pm$ phases of superposition components. These signals are sent to a balanced beam splitter.
The probability of a coincidence count between the detectors monitoring the beam splitter outputs reads \cite{Loudon}
\begin{equation}
P_c = \frac{1}{2} (1-|{\cal V}|^2),
\label{Eq:Pcdef}
\end{equation}
where ${\cal V}$ is given by Eq.~(\ref{Eq:VisibilityDef}) as before. For completeness the above expression has been derived in Appendix. The coincidence count refers to a pair of clicks on both detectors monitoring the output ports $a$ and $b$ of the beam splitter over the timespan of the entire pulse sequence.
A coincidence does not necessarily need to occur within a single time bin. For identical sequences the two photons mode match perfectly and $P_c=0$. Therefore an observation of a coincidence event between the two detectors is an unambiguous signature that the input strings were different, i.e.\ ${\tt x} \neq {\tt y}$. After $N_2$ pairs have been detected by the Referee, the probability that no such event will occur for different strings is $(1-P_c)^{N_2}=[(1+|{\cal V}|^2)/2]^{N_2}$. This expression exhibits exponential dependence in $N_2$, which specifies the number of photons received by the Referee from each party. Hence the scaling in the photon number is analogous to the coherent state protocol, which suggests that the advantage compared to the classical scenario will be similar to the case of first-order interference.
One distinguishing feature is that because two-photon interference is insensitive to the global phase, suppression of coincidence counts will occur also when ${\cal V} = -1$. Consequently, too many differences between individual symbols in error correcting codewords ${\tt E}({\tt x})$ and ${\tt E}({\tt y})$ will have a detrimental effect on the performance of the two-photon protocol. This issue will be carefully analyzed in the next section.
\section{Error correcting code}
\label{Sec:ECC}

The error correcting code is used in the quantum fingerprinting protocol to introduce a minimum gap between the perfect 100\% visibility corresponding to identical input strings and the non-unit visibility for any pair of different strings. For a given error correcting code ${\tt E}$ that maps $n$-bit input strings onto $m$-bit sequences used to modulate the $\pm$ phases of individual optical pulses the visibility defined in Eq.~(\ref{Eq:VisibilityDef}) can be written as:
\begin{equation}
{\cal V}=\frac{1}{m}\bigl[m-2d\bigl({\tt E}({\tt x}),{\tt E}({\tt y})\bigr)\bigr] =1-2\delta.\label{eq:Visibility}
\end{equation}
where $d\bigl({\tt E}({\tt x}),{\tt E}({\tt y})\bigr)$ is the Hamming distance between strings ${\tt E}({\tt x})$ and ${\tt E}({\tt y})$, i.e.\ the number of positions at which the two strings differ, and $\delta = d\bigl({\tt E}({\tt x}),{\tt E}({\tt y})\bigr)/m$ is the relative Hamming distance. For notational simplicity we will omit the argument of $\delta$.
The error correcting code ensures that for any pair of different strings the relative Hamming distance $\delta$ is always greater or equal to a certain minimum value $\delta_{\text{min}}$, i.e. $\delta \ge \delta_{\text{min}}$. The corresponding change in the sequence length is conventionally characterized by the code rate $r = n/m$, which for long input strings satisfies the asymptotic Gilbert-Varshamov bound
$r \le r_{\text{GV}}(\delta_{\text{min}}) = 1-H_{2}(\delta_{\textrm{min}})$, where
$H_{2}(x)=-x\log_{2}x-(1-x)\log_{2}(1-x)$ is the binary entropy function \cite{Lint}. The bound is valid for $\delta_{\textrm{min}} < 1/2$.

\begin{figure}
\includegraphics[scale=1.1]{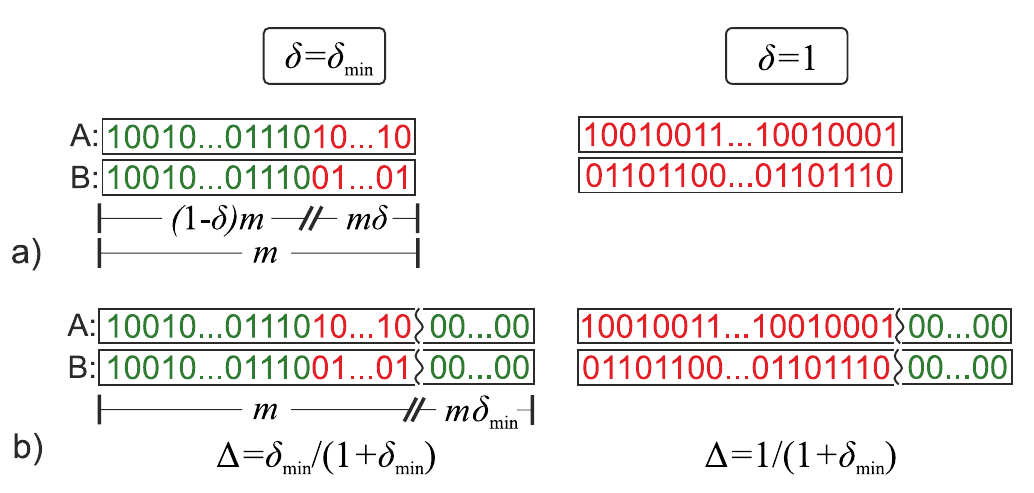}\centering
\caption{Modification of the error correcting code required for the two-photon quantum fingerprinting protocol.
(a) A standard error correcting code of length $m$ guarantees that the relative Hamming distance $\delta$ between any two different codewords used by Alice A and Bob B is between $\delta_{\text{min}}$ and $1$. Two extreme cases are shown. (b) Extension of the standard code by a fixed number of $m\delta_{\text{min}}$ ensures that the relative Hamming distance $\Delta$ for any two codewords in the modified code is bounded between $\delta_{\textrm{min}}/(1+\delta_{\textrm{min}})$ and $1/(1+\delta_{\textrm{min}})$.
\label{fig:Hamming_distance}}
\end{figure}

 The key difference between the coherent-state and the two-photon protocol is that in the latter scenario the relative Hamming distance $\delta$ approaching one, i.e.\ when ${\tt E}({\tt x})$ is a bitwise negation of ${\tt E}({\tt y})$, generates the same outcome as for $\delta=0$ when the sequences are identical. This reflects the insensitivity of two-photon interference to the global phase between the interfering fields. A remedy to this issue is to extend a standard error correcting code by an additional sequence of identical bits, as shown in Fig.~\ref{fig:Hamming_distance}. Specifically, suppose that the initial code of length $m$ is characterized by the minimum relative distance $\delta_{\text{min}}$. If we add extra $m\delta_{\text{min}} $ identical bits to each codeword, the total length of the new code created that way is $M= m(1+\delta_{\text{min}})$, while the relative Hamming distance $\Delta$ between any pair of different codewords after extension satisfies $\Delta_{\text{min}} \le \Delta \le  1-\Delta_{\text{min}}$, where $\Delta_{\text{min}} = \delta_{\text{min}}/(1+\delta_{\text{min}})$. The maximum rate of such a modified code implied by the Gilbert-Varshamov bound is $n/M  \le r_{\text{GV}}(\delta_{\text{min}})/(1+\delta_{\text{min}})$. The right hand side expressed in terms of $\Delta_{\text{min}}$ gives:
\begin{equation}
 \frac{n}{M} = R  \le R_{\text{GV}} (\Delta_{\text{min}}) = (1-\Delta_{\text{min}}) r_{\text{GV}} \left( \frac{\Delta_{\text{min}}}{1-\Delta_{\text{min}}}\right)
\label{Eq:RgeRGV}
\end{equation}
Note that the condition $\delta_{\text{min}} < 1/2$ implies that $\Delta_{\text{min}} < 1/3$.

The bound (\ref{Eq:RgeRGV}) enables one to compare the performance of the coherent state and the two-photon protocols.
In the coherent-state scenario, both Alice and Bob send on average $\bar{n}$ photons. Suppose that the coherent-state protocol utilises a standard error correcting code with a minimum Hamming distance $\delta_{\text{min}}^{\text{coh}}$ which gives maximum visibility ${\cal V}^{\text{coh}} = 1 - 2\delta_{\text{min}}^{\text{coh}}$. The probability that different input strings will not generate any count in the beam splitter output port $b$ is equal to $\exp[-\bar{n}(1-\text{Re} {\cal V}^{\text{coh}})] =\exp(-2 \bar{n} \delta_{\text{min}}^{\text{coh}})$. In the two-photon protocol, Alice and Bob each send $N_2$ photons and need to use a modified error correcting code characterized by a Hamming distance $\Delta_{\text{min}}$. The probability that after receiving $N_2$ photon pairs no coincidence event has occurred is equal to $[(1+|{\cal V}|^2)/2]^{N_2}= \exp\bigl(-N_2 \log[2/(1+|1-2\Delta_{\text{min}}|^2)] \bigr)$, where we have used the visibility for the modified code ${\cal V} = 1-2\Delta_{\text{min}}$.

\begin{figure}[bt]
\includegraphics[scale=0.6]{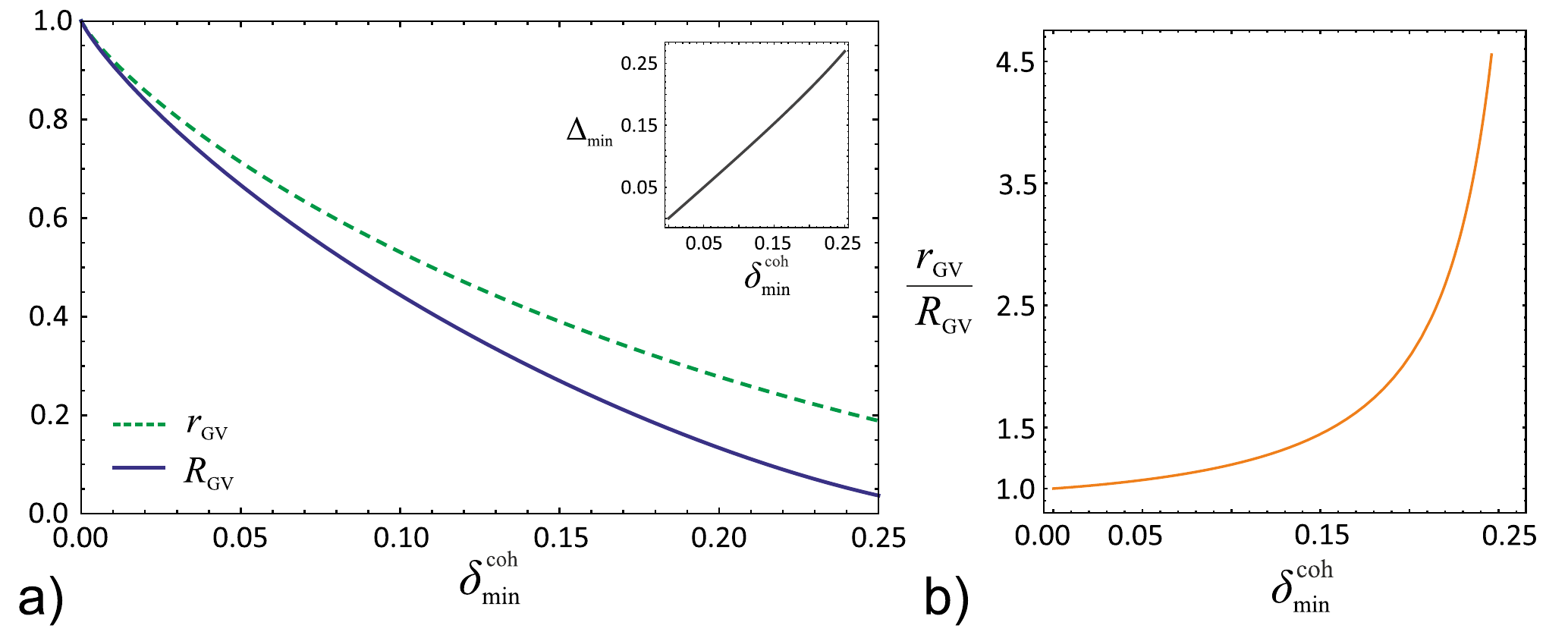}\centering\caption{(a) Maximum code rates $r_{\text{GV}}$  and $R_{\text{GV}}$ given by the Gilbert-Varshamov bound respectively for the coherent-state and the two-photon fingerprinting protocol. The minimum Hamming distances  $\delta_{\text{min}}^{\text{coh}}$ and $\Delta_{\text{min}}$ for the two scenarios are chosen such that the misidentification probability has the same scaling with the number of transmitted photons and their dependence is shown in the inset. (b) The ratio of maximum code rates $r_{\text{GV}}/R_{\text{GV}}$ which specifies the overhead in the codeword length for the two-photon protocol.\label{fig:GV}}
\end{figure}

In order to make a fair comparison between the coherent-state and the two-photon protocol we will assume that the number of photons sent by both Alice and Bob to the Referee is in each scenario similar, i.e. $\bar{n} \approx N_2$, and  that the probability of misidentifying different strings is the same. These requirements give rise to the relation:
\begin{equation}
\Delta_{\text{min}} = \frac{1}{2} [1-\sqrt{2\exp({-2\delta_{\text{min}}^{\text{coh}}})-1}]
\label{Eq:deltamincoh=Delta}
\end{equation}
between the minimum Hamming distances of codes used in both protocols for which the misidentification probabilities are equal. This relation enables a direct comparison of respective code rates. In Fig.~\ref{fig:GV}(a) we depict the Gilbert-Varshamov bound for the error-correcting code used in the coherent-state protocol $r_{\text{GV}}(\delta_{\text{min}}^{\text{coh}})$ along with the bound for the modified code $R_{\text{GV}}(\Delta_{\text{min}})$ employed the in two-photon protocol. In order to ensure the same misidentification probability for both protocols, $\Delta_{\text{min}}$ has been expressed by $\delta_{\text{min}}^{\text{coh}}$ using Eq.~(\ref{Eq:deltamincoh=Delta}). This dependence is shown in the inset of Fig.~\ref{fig:GV}(a).

In order to quantify the increase of the sequence length $M$ for the two-photon protocol compared to the value $m$ in the coherent-state case, in Fig.~\ref{fig:GV}(b) we present the ratio $r_{\text{GV}}(\delta_{\text{min}}^{\text{coh}})/ R_{\text{GV}}(\Delta_{\text{min}})$. Assuming that error correcting codes approaching the Gilbert-Varshamov bound are available, it is seen that for $\delta_{\text{min}}^{\text{coh}} \le 0.25$ the overhead remains approximately below $M/m  \lesssim 5.1$. Let us recall that the classical information capacity of the optical system sent to the Referee by either Alice or Bob in the coherent-state protocol scales as $\bar{n}\log_2 m$ \cite{Luthenhaus2014}. The counterpart for the two-photon protocol is $N_2 \log_2 M$. As long as the minimum Hamming distance remains in the range considered here, these two expressions differ by a term proportional to the number of photons received by the Referee. Consequently, the advantage in the scaling of the quantum protocol based on two-photon interference should be analogous to the coherent-state case.

The above argument can be put on a rigorous footing by calculating the amount of classical information that could be transmitted using physical systems employed in the quantum fingerprinting protocol. This figure needs to be compared with the amount of information that needs to be communicated in  classical fingerprinting. The best known bound on the latter quantity expressed in bits is $I_{\textrm{class}}=(1-2\sqrt{P_{\text{err}}})\sqrt{\frac{n}{2\log 2}}-1$ \cite{Guan}, where $P_{\text{err}}$ is the error probability. Note, however, that it is not known if this bound is saturable and for the best known classical protocol it is necessary to transmit more bits \cite{Guan}. In the quantum two-photon scenario, the number of bits that can be transmitted using at most $N_2$ photons in $MN_2$ modes is equal to $I_{S}=\frac{N_2+1}{M N_2}\log_2{N_2+M N_2 \choose N_2+1}$ \cite{Caves}. This expression takes into account also messages composed from fewer than $N_2$ photons, providing a very conservative figure for the comparison.
In the case of the quantum fingerprinting protocol with coherent states, we will estimate the amount of transmittable classical information $I_{\text{coh}}$ by the capacity of a bosonic channel with the mean photon number $\bar{n}/m$ per channel use, which gives $I_{\text{coh}}=(\bar{n}+m)\log_2(\bar{n}+m)-\bar{n}\log_2\bar{n}-m\log_2 m$ for the entire sequence of $m$ time bins.

\begin{figure}[bt]
\includegraphics[scale=0.4]{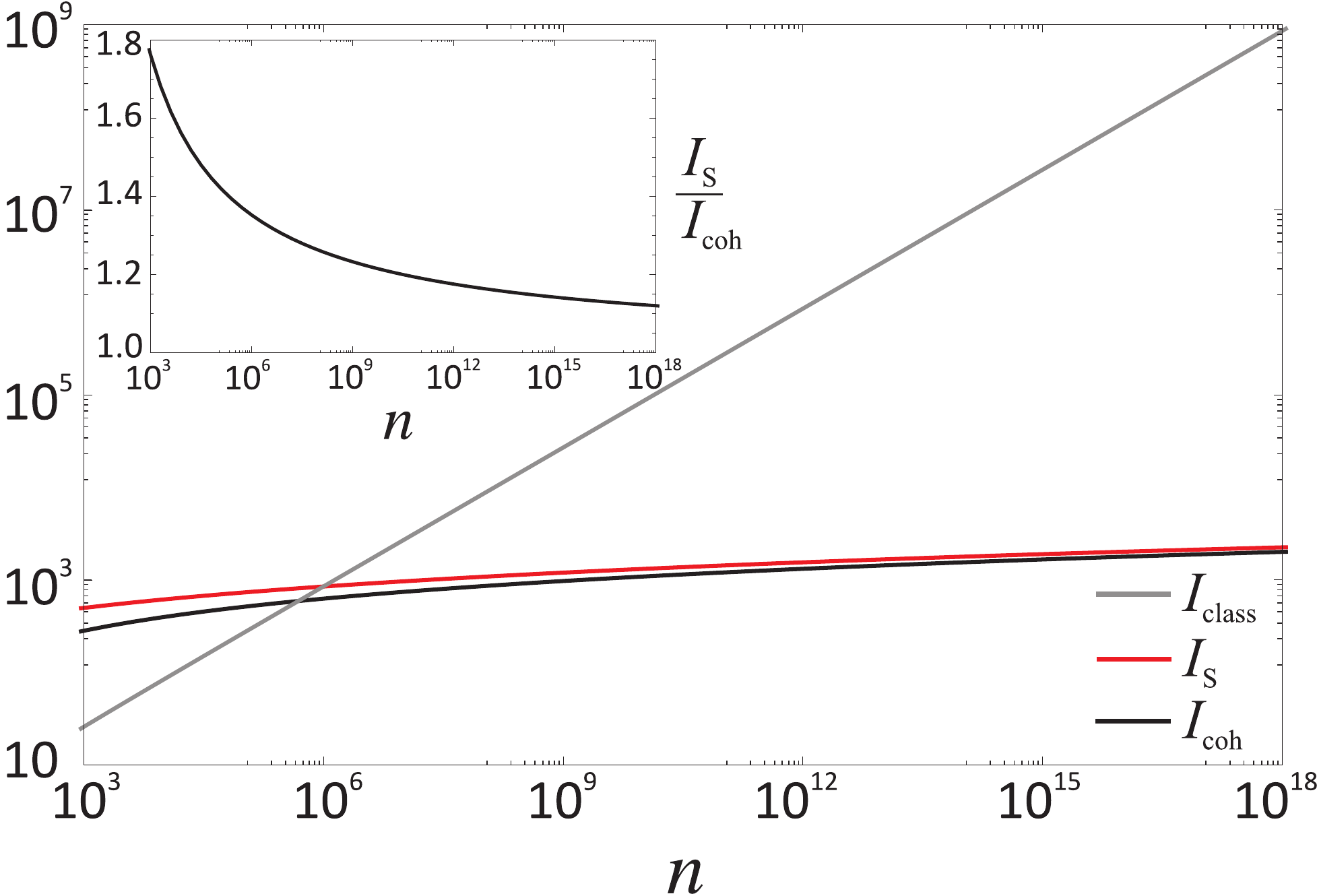}\centering\caption{Comparison of the performance of classical and quantum fingerprinting protocols as a function of the input bit string length $n$. The gray curve represents the best known lower bound on the number of bits required in the classical protocol $I_{\textrm{class}}$. The red and black curves depict the amount of classical information $I_{S}$ and $I_{\text{coh}}$ that could be transmitted in physical systems employed respectively in the two-photon and the coherent state quantum fingerprinting protocol. The error probability is $P_{\text{err}}=10^{-6}$ and the minimum relative Hamming distance equals $\delta^{\text{coh}}_{\text{min}}=0.2$ for the coherent state protocol with the corresponding value for the two-photon protocol given by Eq.~(\ref{Eq:deltamincoh=Delta}). The inset depicts the ratio $I_{S}/I_{\text{coh}}$.\label{fig:information}}
\end{figure}

In Fig.~\ref{fig:information} we compare $I_{\textrm{class}}$, $I_{\textrm{S}}$, and $I_{\text{coh}}$ for the error probability $P_{\text{err}}=10^{-6}$, and the minimum relative Hamming distance $\delta^{\text{coh}}_{\text{min}}=0.2$ for the coherent state protocol, and the corresponding value given by Eq.~(\ref{Eq:deltamincoh=Delta}) for the two-photon protocol. The hypotheses of equal and different input strings are taken as equiprobable with the worst-case scenario, i.e.\ minimum Hamming distance, assumed for the latter. It is seen that the quantum advantage appears for input strings longer than approx.\ $10^{6}$ bits and that the performances of the coherent-state and the two-photon protocols are very similar even though the latter does not make use of a phase reference. Remarkably, as illustrated in the inset in Fig.~\ref{fig:information}, the amount of transmittable classical information for both quantum protocols is nearly the same for large $n$. It can be shown that in this regime both $I_{\textrm{S}}$ and $I_{\text{coh}}$ scale as $O(\log_2 n)$, whereas the bound on classical fingerprinting exhibits a different, square-root scaling $I_{\text{class}}\sim O(\sqrt{n})$.

\section{Realistic two-photon fingerprinting}
\label{Sec:realistic}

In realistic implementations of the quantum fingerprinting protocol one needs to
take into account a variety of experimental imperfections. We will consider a symmetric scenario, in which Alice's and Bob's light sources and optical channels to the Referee have identical properties. The transmission of the channels will be denoted by $\eta$. This parameter can also include the non-unit efficiency of detectors used by the Referee. Further, let the detectors have photon number resolution and exhibit dark counts described by Poissonian statistics with the mean $p_d$ for each detector. In order to describe imperfections of single photon sources used by Alice and Bob, we will take the probability of generating a single photon to be $\bar{n}$. Finally, non-ideal indistinguishability between photons arriving from Alice and Bob will be characterized with a parameter $W$ specifying the fraction of events when the photons interfere perfectly with each other, while in the remaining fraction $1-W$ of cases the photons behave as distinguishable particles.

Assuming the model described above, the Referee receives from Alice the one-photon state $\ket[A]{1_{\tt x}}$ with probability $\eta\bar{n}$. We will also include the possibility of receving simultaneously two photons in a state $\ket[A]{2_{\tt x}} = (\hat{a}^\dagger_{\tt x})^2 /\sqrt{2} \ket[A]{0}$ with a probability denoted as $(\eta\bar{n})^2 g^{(2)}/2$. The same probabilities are associated with the states $\ket[B]{1_{\tt y}}$ and $\ket[B]{2_{\tt y}}=(\hat{b}^\dagger_{\tt y})^2 /\sqrt{2} \ket[B]{0}$ received from Bob. We will assume that two-photon states appear much more rarely compared to one-photon states, i.e.\ $\eta\bar{n} g^{(2)} \ll 1$. The factor $g^{(2)}$ has the interpretation of the normalized second-order intensity correlation function \cite{Mandel} and it allows us to describe imperfections of the single-photon source when $g^{(2)} \ll 1$ as well as include the case of strongly attenuated coherent light sources with a random global phase, when $\eta\bar{n} \ll 1$ and  $g^{(2)}=1$. Additionally, we shall assume that the contribution of dark counts is small compared to genuine photocounts, i.e.\ $p_d / (\eta\bar{n}) \ll 1$.  The overall probability of a two-click event, including both coincidences between the detectors and double counts on one of the detectors, reads:
\begin{equation}
P_{2}=(\eta\bar{n})^{2}\left[1+g^{(2)}+4\frac{p_{d}}{\eta\bar{n}}+2\left(\frac{p_{d}}{\eta\bar{n}}\right)^{2}\right],
\label{eq:doublecountprob}
\end{equation}
where we took into account only the leading terms proportional to $(\eta\bar{n})^2$. A straightforward calculation yields the relative probability of a coincidence event among all two-click events in the form:
\begin{equation}
 Q=\frac{P_c}{P_2} = \frac{1}{2} (1- V_{\text{eff}}), \qquad V_{\text{eff}} = \frac{W|{\cal V}|^2}{1 + g^{(2)} + 4 p_d/(\eta\bar{n}) + 2 p_d^2/(\eta\bar{n})^2}.
 \label{eq:delta2q}
\end{equation}
The remaining fraction $1-Q$ of two-click events corresponds to double counts on one of the detectors placed after the beam splitter.

\begin{figure}
\includegraphics[width=1\textwidth]{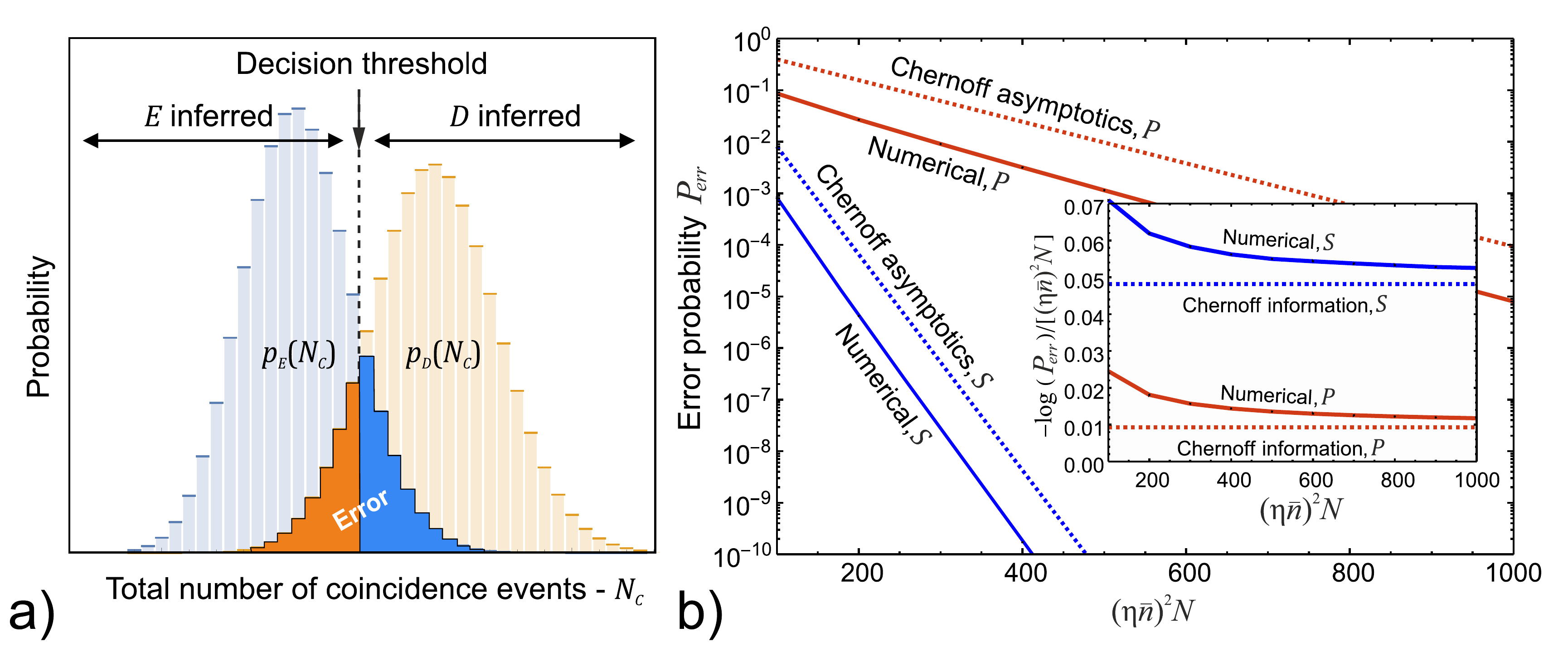}\centering\caption{(a) The decision between hypotheses of different $D$ and equal $E$ inputs is made based on the observed number $N_c$ of coincidence events. The chosen hypothesis corresponds to the higher probability $p_D(N_c)$ or $p_E(N_c)$. (b) The error probability $P_{\text{err}}$ assuming equiprobable hypotheses for the minimum Hamming distance $\Delta_{\mathrm{min}}=0.1$, dark count contribution $p_d/(\eta\bar{n}) = 0.01$, and indistinguishability $W=98\%$. Results for single photons ($S$, blue lines) corresponding to $g^{(2)}=0$ are compared to signals with Poissonian photon number statistics ($P$, red lines). Solid lines are numerical results and dotted lines are asymptotic expressions based on Chernoff information. The plots are parameterized with the effective number $(\eta\bar{n})^2 N$ of pairs composed of photons received from both Alice and Bob. Note that the curves depicting numerical results lie below the lines corresponding to the asymptotic exponential expression. This is because of additional sublinear terms in $N$ in the exponent of the exact error probability.\label{fig:Chernoff_method}}
\end{figure}

Deciding whether the received sequences are different or equal can be viewed as a test between two hypotheses \cite{Gallager}, which we will denote with respective indices $D$ and $E$. These two hypotheses define the corresponding fractions of coincidence events $Q_D$ and $Q_E$ calculated using Eq.~(\ref{eq:delta2q}) respectively for ${\cal V} = 1 - 2 \Delta_{\text{min}}$ and ${\cal V} =1$, assuming the worst-case scenario for the hypothesis $D$. The basis for the decision is the number $N_c$ of coincidence events observed in the overall sample of $N_2$ two-click events. The probability of obtaining a specific $N_c$ is given by a binomial distribution
\begin{equation}
p_{\upsilon}(N_{c})={N_2 \choose N_{c}}Q_{\upsilon}^{N_{c}}(1-Q_{\upsilon})^{N_{2}-N_{c}}, \qquad \upsilon = D, E\label{eq:binomialdistribution}
\end{equation}
Provided that both hypotheses $D$ and $E$ are equiprobable, the average error probability $P_{\text{err}}$ is minimized by a decision rule assigning to each $N_c$ the hypothesis $\upsilon$ for which $p_{\upsilon}(N_{c})$ is larger as shown in Fig.~\ref{fig:Chernoff_method}(a). This yields
\begin{equation}
P_{\text{err}}=\frac{1}{2}\sum_{N_{c}=0}^{N_{2}}\min \{p_{D}(N_{c}),p_{E}(N_{c})\}.\label{eq:numericalerror}
\end{equation}
In Fig.~\ref{fig:Chernoff_method}(b) we depict the error probability for $\Delta_{\text{min}} = 0.1$, $p_d/(\eta\bar{n}) = 0.01$, and $W=0.98$. The single photon case corresponding to $g^{(2)} = 0$ is compared with strongly attenuated phase-averaged coherent states for which $g^{(2)} = 1$. The total number of two-click events is taken as the average value $N_2 \approx P_2 N$ rounded to the nearest integer, where $P_2$ is given by Eq.~(\ref{eq:doublecountprob}) and $N$ is the total number of repetitions of the protocol. In such settings, $(\eta\bar{n})^2 N$ of events has been on average generated by a pair composed of photons received from both Alice and Bob. This figure has been used to parameterize the abscissa of Fig.~\ref{fig:Chernoff_method}(b), which allows one to compare the performance of different light sources used by Alice and Bob with the same photon flux. It is seen that single photon sources perform substantially better compared to classical Poissonian-statistics signals with the same brightness. Note that assumptions made in the calculations imply that the mean photon number received by the Referee from each party in a single run is $\eta\bar{n} \ll 1$.

\section{Chernoff information}
\label{Sec:Chernoff}

In order to quantify the benefit of using non-classical single-photon states we will use the Chernoff information $C$ \cite{CoverThomas}, which allows one to write the error probability for a large number $N$ of repetitions in the asymptotic form
$
P_{\textrm{err}}\sim \exp\left(-NC\right)
$.
Importantly, Chernoff information gives only the leading order of the exponent in $N$ and the exact exponent may also include terms that are sublinear in $N$. We will see in numerical examples that the actual error probability is in general below the asymptotic exponential expression, $P_{\textrm{err}}\leq \exp\left(-NC\right)$. If a single experimental run yields the value of a certain variable $k$ governed by one of two probability distributions $p_D(k)$ or $p_E(k)$, Chernoff information is given explicitly by
\begin{equation}
C=-\log\left[\min_{0 \le \alpha \le 1}\left(\sum_{k} [p_D(k)]^{\alpha}[p_E(k)]^{1-\alpha}\right)\right],\label{eq:chernoff_info}
\end{equation}
where the sum is taken over all possible values of $k$. We will consider hypothesis testing based on coincidence events and double counts on one detector, whose probabilities are given respectively by $Q_\upsilon P_2$ and $(1-Q_\upsilon) P_2$, $\upsilon = D,E$. If no other events are used for inference, the general expression (\ref{eq:chernoff_info}) for Chernoff information can be simplified to
\begin{equation}
C=-\log\left(1-P_{2} + P_2\min_{0 \le \alpha \le 1}[Q_{D}^{\alpha}Q_{E}^{1-\alpha}+(1-Q_{D})^{\alpha}(1-Q_{E})^{1-\alpha}]\right).
\end{equation}
When the overall probability of a two-click event is small, $P_{2}\ll1$, it is justified to expand the above expression  up to the linear term in $P_2$ which gives $ C \approx (\eta\bar{n})^2 \zeta$, where
\begin{equation}
\zeta=\left[1+g^{(2)}+4\frac{p_{d}}{\eta\bar{n}}+2\left(\frac{p_{d}}{\eta\bar{n}}\right)^{2}\right]
\left(1-\min_{0 \le \alpha \le 1}[Q_{D}^{\alpha}Q_{E}^{1-\alpha}+(1-Q_{D})^{\alpha}(1-Q_{E})^{1-\alpha}]\right).\label{eq:dzeta}
\end{equation}
This yields the asymptotic expression for the error probability in the form
\begin{equation}
P_{\text{err}}\sim\exp[-(\eta\bar{n})^{2}N \zeta ]
\label{eq:erroretanbar}
\end{equation}
depicted along numerical results in Fig.~\ref{fig:Chernoff_method}(b). In order to remove the effects of sublinear terms in the exponent of Eq.~(\ref{eq:erroretanbar}), in the inset of Fig.~\ref{fig:Chernoff_method}(b) we compare numerically calculated $-(\log P_{\text{err}})/(\eta\bar{n})^{2}$ with $\zeta$, which shows good asymptotic convergence.

\begin{figure}
\includegraphics[scale=1.1]{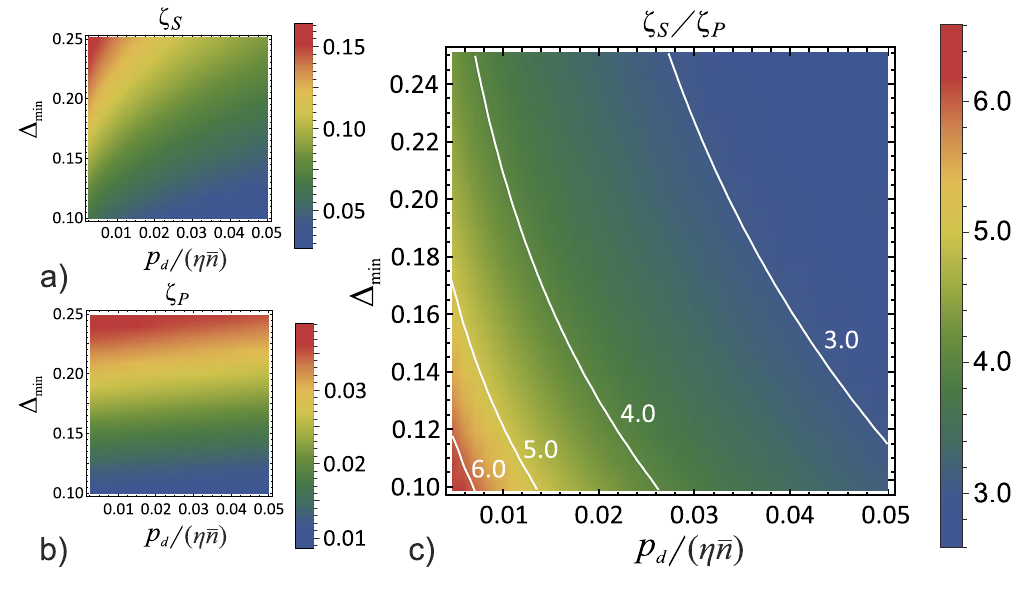}\centering\caption{Rescaled Chernoff information for the fingerprinting protocol based on second-order interference using (a) single photons $\zeta_S$ and (b) optical signals with Poissonian photon number statistics $\zeta_P$ as a function of the dark count contribution $p_d/(\eta\bar{n})$ and the minimum Hamming distance $\Delta_{\text{min}}$. Note different colour codings of numerical values in the panels. (c) The ratio $\zeta_{S}/\zeta_{P}$ which quantifies the benefit of using single photons in the weak signal regime. \label{fig:Rescaled-Chernoff-information}}
\end{figure}

As discussed earlier, the factor $(\eta\bar{n})^{2}N$ specifies the effective number of events generated by pairs composed of photons received from both Alice and Bob. Hence $\zeta$ is the effective Chernoff information per one such event. In Fig.~\ref{fig:Rescaled-Chernoff-information}(a,b) we depict the rescaled Chernoff information $\zeta$ for  light sources with Poissonian statistics characterized by $g^{(2)}=1$ and single photon sources for which $g^{(2)}=0$. We have taken the indistinguishability parameter $W=0.98$ and a rather broad range for the dark count contribution, $0 \le p_d/(\eta\bar{n}) \le 0.05$. Calculations have been performed for the minimum relative Hamming distance $\Delta_{\text{min}}$  between $0.1$ and $0.25$. This parameter enters Eq.~(\ref{eq:delta2q}) through ${\cal V} = 1-2\Delta_{\text{min}}$ when calculating the fraction $Q_D$ of coincidence events for the hypothesis $D$ of different input strings. It is seen that generally the rescaled Chernoff information for single photons $\zeta_S$ is higher than for the Poissonian light $\zeta_P$. The enhancement can be attributed to higher visibility of two-photon interference reaching $100\%$ for general quantum states of light while being limited to $50\%$ for statistical mixtures of coherent states \cite{Wodkiewicz}. This limit can be identified in the expression for the effective visibility $V_{\text{eff}}$ in Eq.~(\ref{eq:delta2q}) by taking the Poissonian statistics with $g^{(2)} = 1$. Other imperfections described by the indistinguishability parameter $W$ and dark count contribution $p_d/(\eta\bar{n})$ only lower further the effective visibility $V_{\text{eff}}$. In Fig.~\ref{fig:Rescaled-Chernoff-information}(c) we show
the ratio $\zeta_S/\zeta_P$, which can be assigned a simple operational interpretation. If we require the same average error probability for the Poissonian and single-photon optical signals with the same photon flux $\eta\bar{n}$, the respective total numbers $N_P$ and $N_S$ of experimental runs satisfy $N_P\zeta_P  = N_S \zeta_S $ according to Eq.~(\ref{eq:erroretanbar}). Hence $\zeta_S/\zeta_P= N_P/N_S$ describes the reduction of the number of experimental runs when single-photons are used in lieu of Poissonian signals with the same brightness. The benefit is most strongly pronounced for lower $\Delta_{\text{min}}$ when different strings in the worst-case scenario only modestly reduce the visibility ${\cal V}$ and when other imperfections, such as dark counts, do not significantly affect the observation of two-photon interference.

\section{Conclusions}
\label{Sec:Conclusions}

We have presented and discussed theoretically a quantum fingerprinting protocol based on two-photon interference which eliminates the need for sharing a phase reference inherent to the coherent-state \cite{Luthenhaus2014,Kumar} and single-particle protocols \cite{Massar} exploiting standard first-order interference. The two-photon protocol requires a modification of the error correcting code to ensure that different strings can generate coincidence events.  We have analyzed the impact of non-Poissonian photon statistics of optical signals and have taken into account a variety of imperfections, such as non-unit channel transmission and detection efficiency, dark counts, and residual distiguishability between interfering photons. The performance of the protocol has been quantified with the help of Chernoff information, which characterizes asymptotic behavior of the error probability.
One should note that in the regime of very weak light sources and/or high channel losses  the rate of useful two-click events scales quadratically as $(\eta\bar{n})^{2}$ with the photon flux $\eta\bar{n}$ which reflects the increasingly rare chance that both the photons from Alice and Bob will be received in the same experimental run. This deficiency could be in principle alleviated by the use of multimode quantum memories at Referee's node \cite{Chrapkiewicz}. It would be also interesting to investigate the effects of phase diffusion \cite{Olivares}, either within the pulse sequences or between the two senders in the case of protocols based on first-order interference.

It is worth pointing out that in the coherent-state quantum fingerprinting protocol detecting clicks with temporal resolution would reveal locations at which individual symbols of the transmitted codewords are effectively compared. A click on a detector monitoring the $b$ port of the beam splitter in a given time bin unambiguously heralds that the respective symbols were different. In the two-photon protocol, a temporally resolved two-click event can be associated with a pair of time bins. The type of the event --- a coincidence or a double count on one detector --- effectively compares the parity of the codeword elements at these two locations for the signals received from Alice and Bob \cite{Rempe}. An analogous observation can be made in two-photon inferference for the spatial degree of freedom \cite{Holo}. Finally, let us note that the photons used by Alice and Bob to imprint codewords as phase sequences should not exhibit temporal correlations which are inherent to e.g.\ photon pair sources based on continuously pumped spontaneous parametric down-conversion. Such correlations could also play the role of shared prior randomness. The photons sent by Alice and Bob should be prepared in the form of narrowband wavepackets that guarantee coherence time over the entire codeword mapped onto the phase pattern. Narrowband photon sources are currently being developed for quantum communication systems that use quantum memories with restricted spectral bandwidth\cite{Fekete,Blatt}.

\section{Acknowledgments}
We wish to acknowledge insightful discussions with N. L\"{u}thenhaus, Q. Zhang and W. Wasilewski.

\section{Funding}
This work is part of the project ``Quantum Optical Communication Systems'' carried out within the TEAM
programme of the Foundation for Polish Science co-financed by the European Union under the European
Regional Development Fund. M.J. was supported by Foundation for Polish Science.

\appendix

\section*{Appendix}

For completeness, we present here detailed derivations of probabilities of events used in Sec.~\ref{Sec:TwoPhotonProtocol} to discuss the performance of the fingerprinting protocols. In the coherent state case, the probability that no count occurs in the output port $b$ is given by the expectation value of a normally ordered exponential expression $\mathop{:}\exp\bigl( - \hat{N}_b \bigr) \mathop{:}$, where $\hat{N}_b$ is the operator of the total number of photons in the output field $b$. Using Eq.~(\ref{Eq:BeamSplitterTransformation}) it can be written in the representation of incoming modes as
\begin{equation}
\hat{N}_b = \frac{1}{2} \sum_{i=0}^{m-1} (\hat{a}_i^\dagger - \hat{b}_{i}^\dagger)(\hat{a}_i - \hat{b}_{i}  ) .
\end{equation}
The expectation value of $\mathop{:}\exp\bigl( - \hat{N}_b \bigr) \mathop{:}$ needs to be evaluated
over the input state $\ket[A]{\alpha_{\tt x}}\ket[B]{\beta_{\tt y}}$ defined in Eq.~(\ref{Eq:InputState}). The expressions $\exp(\alpha\hat{a}_{\tt x}^\dagger - \alpha^\ast \hat{a}_{\tt x})$ and $\exp(\beta\hat{b}_{\tt y}^\dagger - \beta^\ast \hat{b}_{\tt y})$ appearing in  Eq.~(\ref{Eq:InputState}) can be viewed as multimode displacement operators which transform annihilation operators for individual pulses according to
\begin{align}
[\exp(\alpha\hat{a}_{\tt x}^\dagger - \alpha^\ast \hat{a}_{\tt x})]^\dagger \hat{a}_i \exp(\alpha\hat{a}_{\tt x}^\dagger - \alpha^\ast \hat{a}_{\tt x}) & = \hat{a}_i +  (-1)^{{\tt E}_i({\tt x})} \frac{\alpha}{\sqrt{m}}, \nonumber \\
[\exp(\beta\hat{b}_{\tt y}^\dagger - \beta^\ast \hat{b}_{\tt y})]^\dagger \hat{b}_i \exp(\beta\hat{b}_{\tt y}^\dagger - \beta^\ast \hat{b}_{\tt y}) & = \hat{b}_i +  (-1)^{{\tt E}_i({\tt y})} \frac{\beta}{\sqrt{m}}.
\end{align}
Using these formulas, the probability of a zero-count event on the port $b$ is given in terms of the incoming field operators by:
\begin{multline}
\bra[A]{\alpha_{\tt x}}\bra[B]{\beta_{\tt y}}
\mathop{:}\exp ( -\hat{N}_b )\mathop{:}
\ket[A]{\alpha_{\tt x}}\ket[B]{\beta_{\tt y}} \\
= \exp\left( - \frac{1}{2m} \sum_{i=0}^{m-1} | (-1)^{{\tt E}_i({\tt x})} \alpha - (-1)^{{\tt E}_i({\tt y})} \beta|^2  \right)
= \exp[-\bar{n} (1- \text{Re} {\cal V})],
\end{multline}
where in the last expression we have used the definition of the interference visibility in Eq.~(\ref{Eq:VisibilityDef}) and the assumption that $\alpha= \beta = \sqrt{\bar{n}}$. Hence the probability that at least one count is registered at the output port $b$ reads $1-\exp[-\bar{n} (1- \text{Re} {\cal V})]$.

In the two-photon protocol, the probability of a coincidence between the detectors monitoring the output ports of the beam splitter is given by an expectation value of the operator $\mathop{:} \hat{N}_{a} \hat{N}_{b} \mathop{:} $ assuming that the input state contains at most two photons. Here $\hat{N}_a$ is the operator of the total number of photons in the output beam $a$ which expressed in terms of the input modes reads
\begin{equation}
\hat{N}_a = \frac{1}{2} \sum_{j=0}^{m-1} (\hat{a}_j^\dagger + \hat{b}_{j}^\dagger)(\hat{a}_j + \hat{b}_{j}  ) .
\end{equation}
It is now sufficient to note that for the input state $\ket[A]{1_{\tt x}}\ket[B]{1_{\tt y}}$ one has $\hat{a}_i \hat{a}_j \ket[A]{1_{\tt x}}\ket[B]{1_{\tt y}} = 0 = \hat{b}_i \hat{b}_j \ket[A]{1_{\tt x}}\ket[B]{1_{\tt y}}$ and
$\hat{a}_i \hat{b}_j \ket[A]{1_{\tt x}}\ket[B]{1_{\tt y}} = \frac{1}{m}(-1)^{{\tt E}_i({\tt x}) + {\tt E}_j({\tt y})} \ket[A]{0} \ket[B]{0}$. This allows us to simplify
\begin{equation}
\bra[A]{1_{\tt x}}\bra[B]{1_{\tt y}} \mathop{:} \hat{N}_{a} \hat{N}_{b} \mathop{:} \ket[A]{1_{\tt x}}\ket[B]{1_{\tt y}}
= \frac{1}{4m^2} \sum_{i,j=0}^{m-1} | (-1)^{{\tt E}_i({\tt x}) + {\tt E}_j({\tt y})}
- (-1)^{{\tt E}_j({\tt x}) + {\tt E}_i({\tt y})}|^2 = \frac{1}{2} (1- |{\cal V}|^2).
\end{equation}
The second expression, using the definition of the interference visibility from Eq.~(\ref{Eq:VisibilityDef}), yields Eq.~(\ref{Eq:Pcdef}).

\end{document}